\documentclass{article} 
\usepackage{iclr2023_conference,times}

\usepackage{amsmath,amsfonts,bm}

\def\eqref#1{equation~\ref{#1}}

\def\1{\bm{1}}

\def\mA{{\bm{A}}}

\def\mV{{\bm{V}}}

\DeclareMathAlphabet{\mathsfit}{\encodingdefault}{\sfdefault}{m}{sl}
\SetMathAlphabet{\mathsfit}{bold}{\encodingdefault}{\sfdefault}{bx}{n}
\newcommand{\tens}[1]{\bm{\mathsfit{#1}}}

\def\tI{{\tens{I}}}

\usepackage{hyperref}
\usepackage{url}
\usepackage{multirow}
\usepackage{comment}
\usepackage{amsmath,amssymb}
\usepackage{graphicx}
\usepackage{float}
\usepackage{subfig}
\usepackage{hyperref}

\title{Graph-Based Deep Learning for Sea Surface Temperature Forecasts}

\author{Ding Ning \\
School of Mathematics and Statistics\\
University of Canterbury\\
Christchurch 8041, New Zealand \\
\texttt{ding.ning@pg.canterbury.ac.nz} \\
\AND
Varvara Vetrova \\
School of Mathematics and Statistics\\
University of Canterbury\\
Christchurch 8041, New Zealand \\
\texttt{varvara.vetrova@canterbury.ac.nz} \\
\And
Karin R. Bryan \\
School of Science\\
University of Waikato\\
Hamilton 3216, New Zealand \\
\texttt{karin.bryan@waikato.ac.nz} \\
}

\iclrfinalcopy 
\begin{document}

\maketitle

\begin{abstract}
Sea surface temperature (SST) forecasts help with managing the marine ecosystem and the aquaculture impacted by anthropogenic climate change. Numerical dynamical models are resource intensive for SST forecasts; machine learning (ML) models could reduce high computational requirements and have been in the focus of the research community recently. ML models normally require a large amount of data for training. Environmental data are collected on regularly-spaced grids, so early work mainly used grid-based deep learning (DL) for prediction. However, both grid data and the corresponding DL approaches have inherent problems. As geometric DL has emerged, graphs as a more generalized data structure and graph neural networks (GNNs) have been introduced to the spatiotemporal domains. In this work, we preliminarily explored graph re-sampling and GNNs for global SST forecasts, and GNNs show better one month ahead SST prediction than the persistence model in most oceans in terms of root mean square errors.
\end{abstract}

\section{Introduction}

The variability of SSTs, or SST anomalies, is associated with multiple climate oscillations or extreme events, such as the El Ni\~{n}o-Southern Oscillation (ENSO), the Indian Ocean Dipole (IOD) oscillation, and marine heatwaves. The ability to accurately forecast SST variability would allow mitigation of its potential impact, such as by collecting healthy samples for repopulation of impacted ecosystems and adjusting aquaculture production beforehand.

A number of DL models have been developed to predict SSTs and/or related events. Early work started with convolutional neural networks (CNNs). \cite{ham2019deep,ham2021unified} used a CNN to predict ENSO up to 18 months in advance and \cite{cachay2021world} used a GNN to improve the forecasts for one to six lead months. IOD forecasts have been made using a CNN \citep{fengpredictability} and a long short-term memory (LSTM) network \citep{pravallika2022prediction} respectively. A CNN was developed to forecast SSTs and marine heatwaves around Australia \citep{boschetti2022sea}. Later work started to address the combination of multiple neural network classes for SST forecasts. \cite{taylor2022deep} combined a U-Net \citep{ronneberger2015u} with an LSTM \citep{unetlstm} to forecast global SSTs up to 24 months ahead and validated the forecasts with a focus on specific SST variability-related oscillations (ENSO and IOD) and events (the ``Blob'' marine heatwave). The DL models outlined above input sequences or grids, i.e. Euclidean data, and used image or video processing techniques to perform SST forecasts. However, there is a potential for further improvement via utilizing the structure of climatological data, which are different from images and videos.

Non-Euclidean graphs could be an alternative to grids. Graph representation learning has been successfully applied to domains such as social networks \citep{gupta2021graph} and bioinformatics \citep{yi2022graph}. The teleconnections of climate events \citep{tsonis2006networks}, either through atmosphere, oceanic circulation, or large-scale oceanic waves, are increasingly considered as an important factor for developing DL methods \citep{cachay2021world,taylor2022deep} and could be modeled by graphs. Grids and CNNs still have some inherent problems, such as replacement for missing values, rotation equivariance \citep{defferrard2019deepsphere}, and receptive fields \citep{luo2016understanding}, making them difficult to use in modeling global oceans. Graph-based DL for SST forecasts is not as well explored as the grid-based. Hence, we investigated whether graphs and graph-based DL are suited for SST forecasts. We started by extending the work by \cite{taylor2022deep} to the graph domain and found that GNNs generally outperform the persistence model for one month ahead SST forecasts globally.

\section{Data}

\textbf{Dataset.} The dataset for SST forecasts is from ERA5 \citep{hersbach2020era5}. ERA5 is a reanalysis product that provides monthly estimates of a large number of atmospheric, land and oceanic variables at global scale with a spatial resolution of 0.25°, from 1950 to 1978 (the preliminary version) and from 1959 to the present (the current version).

\textbf{Data Preprocessing.} We downloaded the ERA5 data with the univariate SST from both versions. Two versions of the data were joined along the time axis, using the preliminary version from January 1950 to December 1978 and the current version from January 1979 to August 2022. Following \cite{taylor2022deep}, we used the Climate Data Operators (CDO) \citep{https://doi.org/10.5281/zenodo.3539275} to process the joined dataset to a [64, 128, 872] latitude (64°S to 62°N in 2° increments), longitude (180°W to 180°E in 2.8125° increments), and month (January 1950 to August 2022 in one month increment) grid. The unit of SSTs is Kelvin. We normalized the data to the [-1, 1] range using the following formula:
\begin{align*}
\tilde{x}_i=\frac{x_i-x_{min}}{x_{max}-x_{min}}\cdot2-1,
\end{align*}
where $x_i$ is a raw ERA5 SST value, $x_{min}$ and $x_{max}$ are the minimum and the maximum over all data, and $\tilde{x_i}$ is a normalized SST value, which resulted in a normalized [64, 128, 872] grid. Normalization primarily helps to stabilize numerical calculations and accelerate the rate of convergence to a solution \citep{taylor2022deep}. The first 760 time steps were used for training and the remaining were used for testing. Unlike \cite{taylor2022deep}, we did not use the two-meter atmospheric temperature variable to interpolate the land pixels in the SST grid.

\section{Methods}

\subsection{Graph Construction}

We constructed the graphs by defining the adjacency matrix and the node attribute matrix. We have not found suitable relational variables for SST forecasts, so the edge attribute matrix was left empty.

\textbf{Node Attribute Matrix.} Let $\displaystyle \tI\in\mathbb{R}^{M\times N\times T}$ denote a tensor that represents the preprocessed SST grid, where $M$ is the number of points in the latitudinal direction, $N$ is the number of points in the longitudinal direction, and $T$ is the number of monthly time steps.
There is an SST time series of $T$ elements at every latitude and longitude coordinate. The node attribute matrix $\displaystyle \mV\in\mathbb{R}^{X\times T}$, where $X=M\times N$ is the number of nodes, was acquired by flattening every 2D slice $\displaystyle \tI_{:,:,t}$ of $\displaystyle \tI$ at time step $t$. $V_{x,t}$ is the SST value at the $x$\textsuperscript{th} node at time step $t$.

\textbf{Adjacency Matrix.} We constructed a set of undirected graphs and a set of directed graphs. For the undirected graphs, an element $A$ in the adjacency matrix $\displaystyle \mA$ is defined by an indicator function:
\begin{align*}
A_{x,y}=\displaystyle \1_\mathrm{|\rho(\pmb{V}_{x,:}, \pmb{V}_{y,:})|>c},\\
A_{x,y}=A_{y,x},
\end{align*}
where $\displaystyle \mV_{x,:}$ and $\displaystyle \mV_{y,:}$ are the SST time series at any two nodes, 
$\rho(\cdot)$ is the Pearson correlation coefficient in this case but could be other measures, and $c$ is a threshold as a controllable parameter. For the directed graphs, with regards to one lead time forecasts, when the correlation between the time series at node $x$ and one lead time series at node $y$ is above the threshold, we consider that there is an edge between the two nodes, the source node is $x$ and the destination node is $y$. Therefore, an element $\tilde{A}$ in the adjacency matrix $\tilde{\displaystyle \mA}$ for a directed graph is defined as
\begin{align*}
\tilde{A}_{x,y}=\displaystyle \1_\mathrm{|\rho(\pmb{V}_{x,0:T-1}, \pmb{V}_{y,1:T})|>c}.
\end{align*}

The decrease in the correlation threshold $c$ leads to a substantial increase in the number of edges and node degrees. We generated multiple sets of SST graphs, with the statistics shown in Table \ref{tb:data}. Besides, all graphs have isolated nodes and no self-loops, and graphs in the same set are homogeneous.

These graph data have been made available for download and the link is in the Appendix.

\begin{table}[h]
\caption{Statistics of the SST graphs from ERA5. The average node degree is the average number of edges per node. The sets used to train GNN models are in bold.}\label{tb:data}
\begin{center}
\begin{tabular}{lllll}
\textbf{\textbf{Number of nodes}} & \textbf{\textbf{Is directed}} & \textbf{$\pmb{c}$} & \textbf{Number of edges} & \textbf{Average node degree} \\ \hline
\multirow{7}{*}{\textbf{5774}}    & No                            & NA                 & 0                        & 0                            \\
                                  & No                            & 0.99               & 8090                     & 1.4                          \\
                                  & No                            & 0.97               & 88510                    & 15.33                        \\
                                  & \textbf{No}                   & \textbf{0.95}      & \textbf{325546}          & \textbf{56.38}               \\
                                  & No                            & 0.9                & 2949098                  & 510.75                       \\
                                  & \textbf{Yes}                  & \textbf{0.9}       & \textbf{292260}          & \textbf{50.62}               \\
                                  & Yes                           & 0.8                & 5125450                  & 887.68
\end{tabular}
\end{center}
\end{table}

\subsection{Graph Neural Networks}

We applied widely-used GNN classes to perform learning on SST graphs: graph convolutional networks (GCNs) \citep{kipf2016semi}, graph attention networks (GATs) \citep{velivckovic2017graph}, and GraphSAGE \citep{hamilton2017inductive} for undirected graphs, and relational GCNs (RGCNs) \citep{schlichtkrull2018modeling} for directed graphs. These GNN models were implemented in Python using PyTorch \citep{paszke2019pytorch} and PyTorch Geometric \citep{fey2019fast}.

The forecasting task here is node regression with sliding windows. We aimed at forecasting SSTs at every node one month ahead. In each iteration, the inputs were a set of $w$ graphs at earlier time steps, $\pmb{V}_{:,t-w+1,...,t}$, where $w$ is the forecasting window size, and the output was one graph at the time step for prediction, $\pmb{V}_{:,t+1}$. Following \cite{taylor2022deep}, we used a window size of 12.

We deployed a similar structure for all GNNs: there are two layers, where the first layer inputs $w$ features and outputs 30 features, and the second layer inputs 30 features and outputs 1 feature. The optimizer is the root mean square propagation, with a 0.001 learning rate, a 0.9 alpha, a 0.9 momentum, and a 0.0001 weight decay. The activation is the hyperbolic tangent. The loss is the mean squared error. The root mean squared error (RMSE) is reported. The number of training epochs is 20. For the GAT, the number of heads is eight; for the RGCN, the number of relations is two and the number of bases is four. The GCN, the GAT, and the GraphSAGE were all trained using undirected graphs with $c=0.95$; the RGCN was trained using directed graphs with $c=0.9$. We chose these two values of $c$ as they lead to similar average node degrees. In turn, it allowed us to make use of limited computational resources during the exploratory phase in order to identify an appropriate GNN class. Our future plan is to experiment with different values of $c$ and consequently with graphs of larger size.

\section{Results and Discussion}

One model was trained for each model class. We calculated RMSEs on the test data for each node. The average RMSEs of all nodes are summarized in Table \ref{tb:results} for all GNN models and for the persistence model as a baseline against which to compare performance.

\begin{table}[h]
\caption{Average RMSEs across all nodes. The GNN model that outperforms the persistence model is in bold.}\label{tb:results}
\begin{center}
\begin{tabular}{lll}
\textbf{\textbf{Model class}} & \textbf{\textbf{Average RMSE}} & \textbf{\textbf{Average RMSE in Kelvin}} \\ \hline
Persistence                   & 0.0621                         & 1.1380                                 \\
GCN                           & 0.1135                         & 2.0789                                 \\
GAT                           & 0.1107                         & 2.0269                                 \\
\textbf{GraphSAGE}            & \textbf{0.0346}                & \textbf{0.6338}                        \\
RGCN                          & 0.3506                         & 6.4207
\end{tabular}
\end{center}
\end{table}

\begin{figure}[h]
\centering
  \includegraphics[scale=0.34]{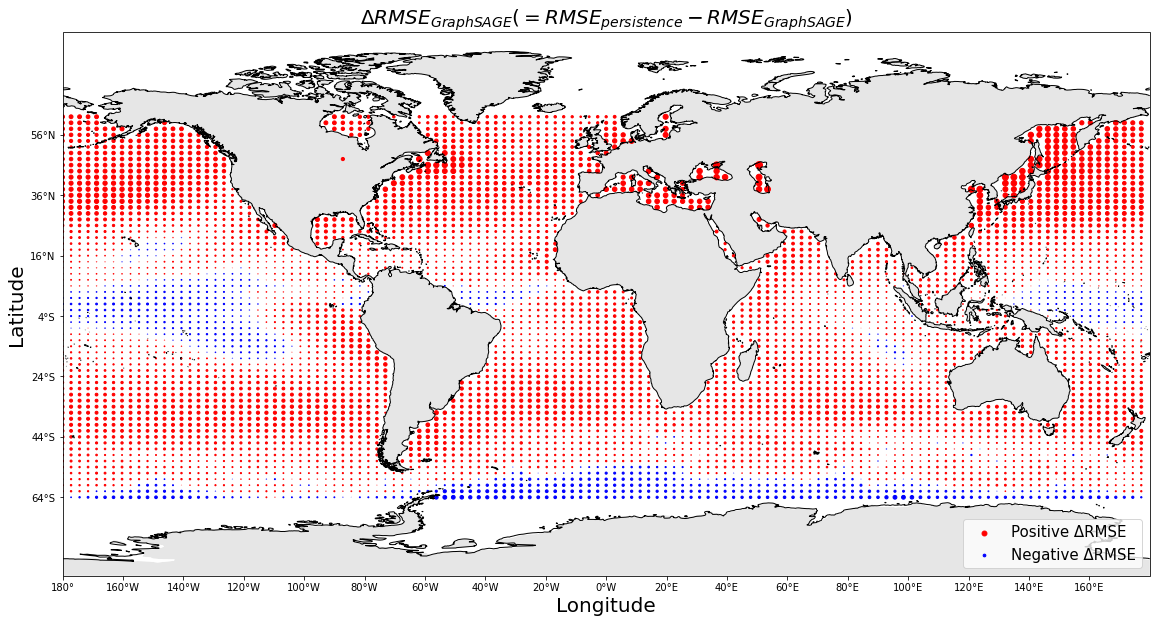}
  \caption{$\Delta RMSE_{GraphSAGE}$($=RMSE_{persistence}-RMSE_{GraphSAGE}$) for the one month ahead prediction at every node on the world map. The positive values are red, indicating the GraphSAGE outperforms the persistence model in terms of the test RMSE at a node, and the negative values are blue; the size of a dot represents the absolute value.}\label{fg:sageres}
\end{figure}

Only the GraphSAGE outperforms the persistence model in terms of average RMSEs. The GCN and the GAT may need further hyperparameter tuning and more complex structures. For the RGCN, the problem might arise with the directed graphs. Additionally, the GraphSAGE took the least amount of time to train, indicating its superior time efficiency when applied to the SST graphs. In order to further investigate the performance of the GraphSAGE, we obtained the difference between the persistence RMSE and the RMSE of the GraphSAGE per node, shown in Figure \ref{fg:sageres}.

The GraphSAGE model generally outperforms persistence across the world, especially in the temperate zones, possibly because the changes in temperate SSTs are stable. The model performs poorly in the tropics, given that the changes in SSTs in the tropics are slight and irregular. The predictions near continents are generally better. The predictions near Antarctica are generally poor.

In the Appendix, we selected some locations from Figure \ref{fg:sageres} in Figure \ref{fg:locs} and created time series plots and scatter plots for the GraphSAGE in Figures \ref{fg:ts} and \ref{fg:sc} respectively. At most locations, the general trends and cycles are predicted. The predictions of minor variations or extreme values could be improved.

\section{Conclusion}

We have examined the potential of GNNs for forecasting global SSTs one month ahead. The GraphSAGE model outperforms the persistence model at most places across the world in terms of RMSEs and  especially provides better predictions in the temperate zones and near continents.

Based on the initial results, we assume that graphs and GNN models can be substitutes for grids and CNN models for global oceanic variable forecasts, with regards to avoiding the problems of grids and CNNs, utilizing the flexibility of graph re-sampling, and the computational efficiency of GNNs.

\subsection{Future Work}

This work constitutes a step towards the forecasting of seasonal SST anomalies and marine heatwaves using graph-based deep learning methods. The following work is suggested.

\textbf{Model Tuning.} For the current GNN models, there is room for improvement by tuning hyperparameters and adding auxiliary techniques. Similar to the U-Net for SST forecasts \citep{taylor2022deep}, graph U-Nets \citep{gao2019graph} could be another GNN class for consideration.

\textbf{Graph Construction.} So far, we have not included edge attributes to reflect GNNs' capability of learning relational variables. Finding these useful oceanic or atmospheric variables will possibly improve the forecasts. In addition, aspects such as selecting non-parametric measures for $\rho(\cdot)$ and removing seasonality would also alter the results. Graph construction from grids is an ongoing problem due to its flexibility and influence on overall performance.

\textbf{Anomaly Prediction.} Forecasting SST anomalies and their associated extreme events is of greater ecological and socioeconomic value. When predicting anomalies, the node regression will be reformulated as a node imbalanced regression task, which requires additional techniques to handle.

\textbf{Long Lead Forecasts.} Accurate long lead SST forecasts will help with planning and taking actions earlier to mitigate the impacts of SST extremes. We are interested in forecasts from three months to two years in advance. \cite{taylor2022deep} have demonstrated that using an autoregressive approach by repeatedly feeding the short lead predictions back to models and adding an LSTM layer make long lead forecasts achievable.

\bibliography{iclr2023_conference}
\bibliographystyle{iclr2023_conference}

\appendix
\section{Appendix}\label{apd}

\textbf{Data Availability.} ERA5 can be downloaded from the Copernicus Climate Change Service, \url{https://cds.climate.copernicus.eu/}. The SST graphs generated from ERA5 are accessible at \url{https://doi.org/10.5281/zenodo.7755727}.\\

\begin{figure}[h]
\centering
  \includegraphics[scale=0.34]{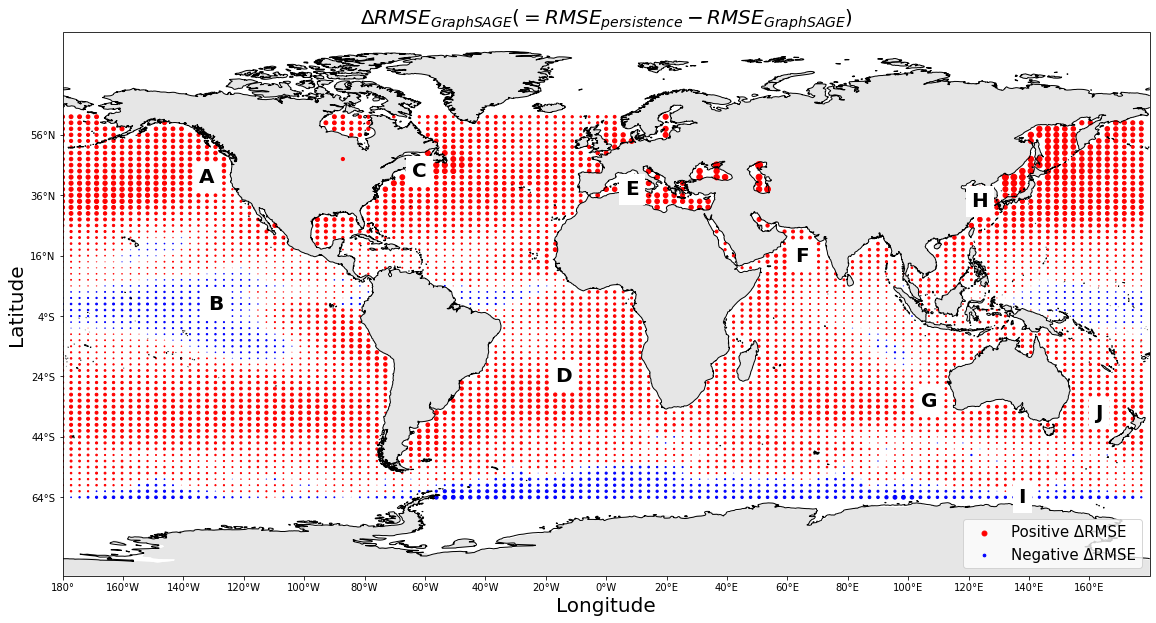}
  \caption{Ten selected locations for time series plots, marked in bold black on the $\Delta RMSE_{GraphSAGE}$ map.}\label{fg:locs}
\end{figure}

\begin{figure}[h]
\centering
\includegraphics[scale=0.46]{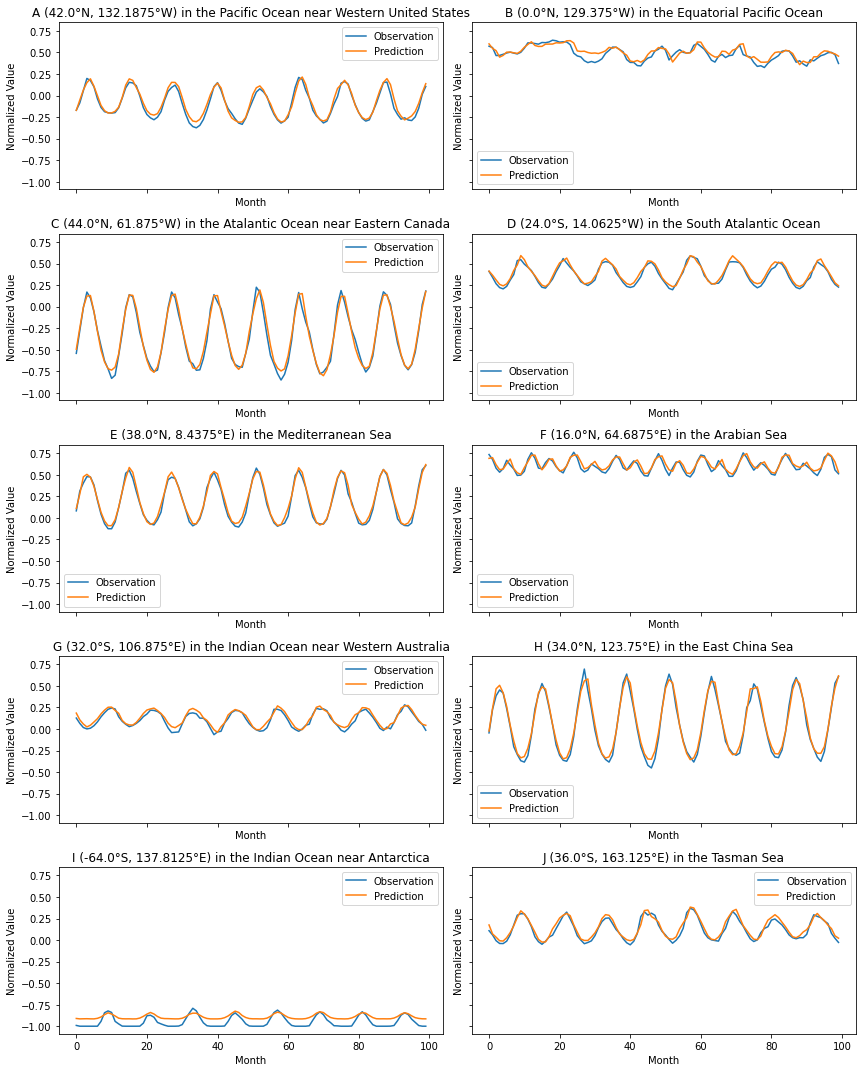}
\caption{Time series plots: predictions by the GraphSAGE and the corresponding observations in the test period at the ten selected locations.}\label{fg:ts}
\end{figure}

\begin{figure}[h]
\centering
\includegraphics[scale=0.43]{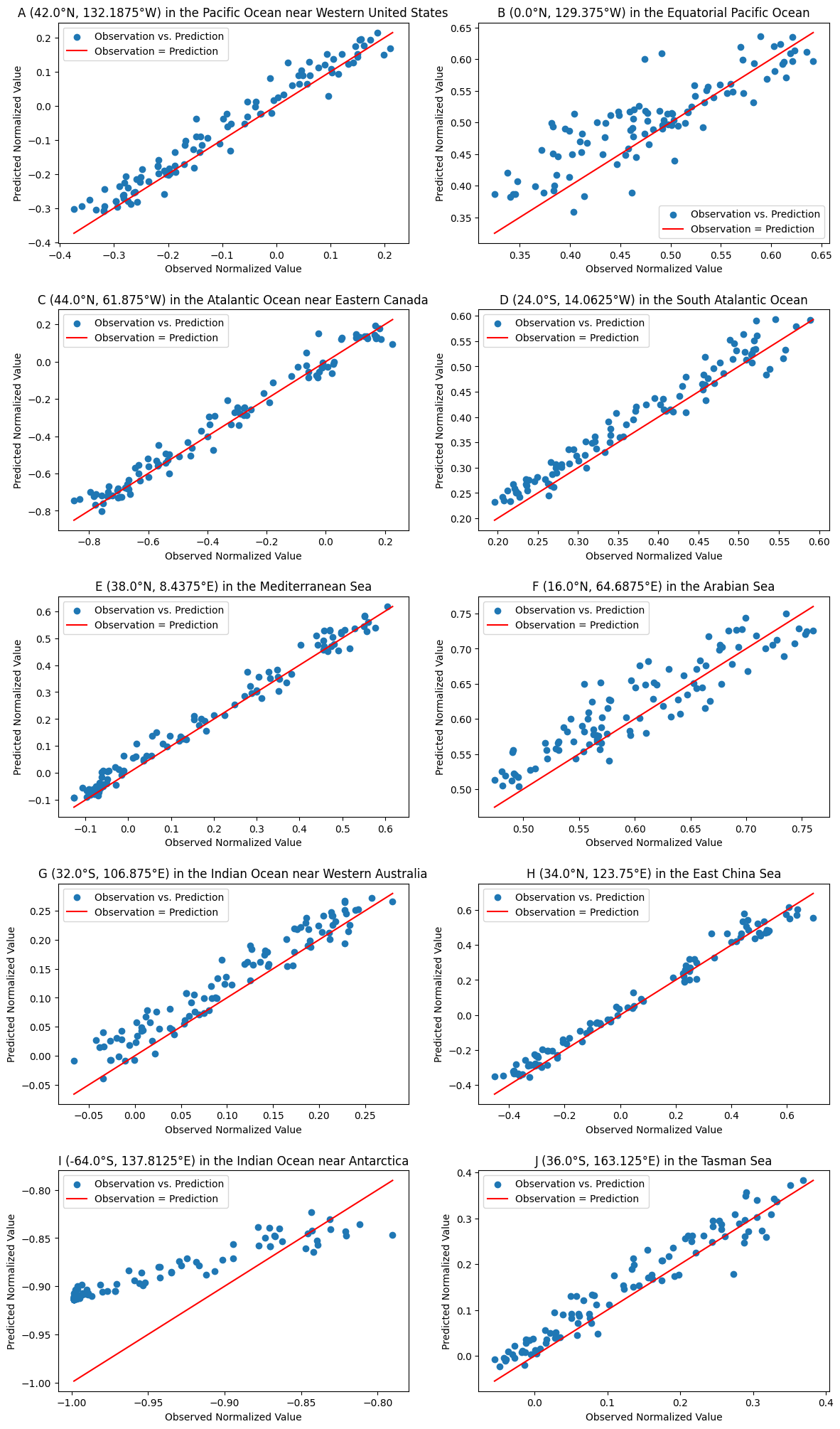}
\caption{Scatter plots: predictions by the GraphSAGE and the corresponding observations in the test period at the ten selected locations.}\label{fg:sc}
\end{figure}

\end{document}